\documentclass[reprint,amsmath,amssymb,aps]{revtex4-1}
\usepackage{bm}
\usepackage{dcolumn}
\usepackage{lineno}
\usepackage{graphicx}
\usepackage{subfigure}
\usepackage{epstopdf}
\usepackage{float}
\usepackage{threeparttable}
\usepackage{overpic}
\usepackage{subfigure}
\usepackage{float}
\usepackage{enumitem}
\usepackage{ulem}
\usepackage{bm}
\usepackage{cmap}
\usepackage{comment}
\usepackage{mathtext}
\usepackage{mathrsfs}
\usepackage{multirow}
\usepackage{xcolor}
\definecolor{aa}{RGB}{0,0,139}
\usepackage[bookmarksnumbered=true,colorlinks,urlcolor=aa,linkcolor=blue,anchorcolor=aa,citecolor=aa]{hyperref}

\begin{document}

\newcommand{\afs}{\alpha_s}
\newcommand{\bgp}{\beta\gamma}
\newcommand{\eff}{\varepsilon}
\newcommand{\sintht}{\sin{\theta}}
\newcommand{\costht}{\cos{\theta}}
\newcommand{\dedx}{dE/dx}

\newcommand{\probfc}{Prob_{\chi^2}}
\newcommand{\probpi}{Prob_{\pi}}
\newcommand{\probka}{Prob_{K}}
\newcommand{\probpr}{Prob_{p}}
\newcommand{\proball}{Prob_{all}}

\newcommand{\chicJ}{\chi_{cJ}}
\newcommand{\gchicJ}{\gamma\chi_{cJ}}
\newcommand{\gchica}{\gamma\chi_{c0}}
\newcommand{\gchicb}{\gamma\chi_{c1}}
\newcommand{\gchicc}{\gamma\chi_{c2}}
\newcommand{\hc}{h_c(^1p_1)}
\newcommand{\qqb}{q\bar{q}}
\newcommand{\uub}{u\bar{u}}
\newcommand{\ddb}{d\bar{d}}
\newcommand{\SSB}{\Sigma^+\bar{\Sigma}^-}
\newcommand{\ccb}{c\bar{c}}

\newcommand{\psipto}{\psi^{\prime}\rightarrow \pi^+\pi^- J/\psi}
\newcommand{\ptomm}{J/\psi\rightarrow \mu^+\mu^-}
\newcommand{\ppp}{\pi^+\pi^- \pi^0}
\newcommand{\pip}{\pi^+}
\newcommand{\pim}{\pi^-}
\newcommand{\kap}{K^+}
\newcommand{\kam}{K^-}
\newcommand{\ks}{K^0_s}
\newcommand{\pbar}{\bar{p}}
\newcommand{\jp}{J/\psi\rightarrow \gamma\pi^0}
\newcommand{\je}{J/\psi\rightarrow \gamma\eta}
\newcommand{\jep}{J/\psi\rightarrow \gamma\eta^{\prime}}

\newcommand{\LL}{\ell^+\ell^-}
\newcommand{\EE}{e^+e^-}
\newcommand{\MM}{\mu^+\mu^-}
\newcommand{\GG}{\gamma\gamma}
\newcommand{\TT}{\tau^+\tau^-}
\newcommand{\pp}{\pi^+\pi^-}
\newcommand{\kk}{K^+K^-}
\newcommand{\ppb}{p\bar{p}}
\newcommand{\gpp}{\gamma \pi^+\pi^-}
\newcommand{\gkk}{\gamma K^+K^-}
\newcommand{\gppb}{\gamma p\bar{p}}
\newcommand{\ggee}{\gamma\gamma e^+e^-}
\newcommand{\gguu}{\gamma\gamma\mu^+\mu^-}
\newcommand{\ggll}{\gamma\gamma l^+l^-}
\newcommand{\ppee}{\pi^+\pi^- e^+e^-}
\newcommand{\ppuu}{\pi^+\pi^-\mu^+\mu^-}
\newcommand{\etap}{\eta^{\prime}}
\newcommand{\gpi}{\gamma\pi^0}
\newcommand{\geta}{\gamma\eta}
\newcommand{\getap}{\gamma\etap}
\newcommand{\pppp}{\pi^+\pi^-\pi^+\pi^-}
\newcommand{\ppkk}{\pi^+\pi^-K^+K^-}
\newcommand{\pppr}{\pi^+\pi^-p\bar{p}}
\newcommand{\kkkk}{K^+K^-K^+K^-}
\newcommand{\kskp}{K^0_s K^+ \pi^- + c.c.}
\newcommand{\ppkp}{\pi^+\pi^-K^+ \pi^- + c.c.}
\newcommand{\ksks}{K^0_s K^0_s}
\newcommand{\dphi}{\phi\phi}
\newcommand{\phikk}{\phi K^+K^-}
\newcommand{\ppeta}{\pi^+\pi^-\eta}
\newcommand{\gpppp}{\gamma \pi^+\pi^-\pi^+\pi^-}
\newcommand{\gppkk}{\gamma \pi^+\pi^-K^+K^-}
\newcommand{\gpppr}{\gamma \pi^+\pi^-p\bar{p}}
\newcommand{\gkkkk}{\gamma K^+K^-K^+K^-}
\newcommand{\gkskp}{\gamma K^0_s K^+ \pi^- + c.c.}
\newcommand{\gppkp}{\gamma \pi^+\pi^-K^+ \pi^- + c.c.}
\newcommand{\gksks}{\gamma K^0_s K^0_s}
\newcommand{\gphiphi}{\gamma \phi\phi}

\newcommand{\tpp}{3(\pi^+\pi^-)}
\newcommand{\tppkk}{2(\pi^+\pi^-)(K^+K^-)}
\newcommand{\pptkk}{(\pi^+\pi^-)2(K^+K^-)}
\newcommand{\tkk}{3(K^+K^-)}
\newcommand{\gtpp}{\gamma 3(\pi^+\pi^-)}
\newcommand{\gtppkk}{\gamma 2(\pi^+\pi^-)(K^+K^-)}
\newcommand{\gpptkk}{\gamma (\pi^+\pi^-)2(K^+K^-)}
\newcommand{\gtkk}{\gamma 3(K^+K^-)}

\newcommand{\psp}{\psi(3686)}
\newcommand{\jpsi}{J/\psi}
\newcommand{\ar}{\rightarrow}
\newcommand{\lra}{\longrightarrow}
\newcommand{\jpsito}{J/\psi \rightarrow }
\newcommand{\ptoppjp}{J/\psi \rightarrow\pi^+\pi^- J/\psi}
\newcommand{\pspto}{\psi^\prime \rightarrow }
\newcommand{\ptop}{\psi'\rightarrow\pi^0 J/\psi}
\newcommand{\ptoeta}{\psi'\rightarrow\eta J/\psi}
\newcommand{\ecto}{\eta_c \rightarrow }
\newcommand{\ecpto}{\eta_c^\prime \rightarrow }
\newcommand{\xto}{X(3594) \rightarrow }
\newcommand{\chicJto}{\chi_{cJ} \rightarrow }
\newcommand{\chiczto}{\chi_{c0} \rightarrow }
\newcommand{\chicoto}{\chi_{c1} \rightarrow }
\newcommand{\chictto}{\chi_{c2} \rightarrow }
\newcommand{\pspp}{\psi^{\prime\prime}}
\newcommand{\ptochic}{\psi(2S)\ar \gamma\chi_{c1,2}}
\newcommand{\ppjpsi}{\pi^0\pi^0 J/\psi}
\newcommand{\utoeta}{\Upsilon^{\prime}\ar\eta\Upsilon}
\newcommand{\ww}{\omega\omega}
\newcommand{\wf}{\omega\phi}
\newcommand{\ff}{\phi\phi}
\newcommand{\npsp}{N_{\psp}}
\newcommand{\llb}{\Lambda\bar{\Lambda}}
\newcommand{\llbpi}{\llb\pi^0}
\newcommand{\llbeta}{\llb\eta}
\newcommand{\ppi}{p\pi^-}
\newcommand{\pbpi}{\bar{p}\pi^+}
\newcommand{\lamb}{\bar{\Lambda}}
\def\ctup#1{$^{\cite{#1}}$}
\newcommand{\bfg}{\begin{figure}}
\newcommand{\efg}{\end{figure}}
\newcommand{\bitm}{\begin{itemize}}
\newcommand{\eitm}{\end{itemize}}
\newcommand{\bnum}{\begin{enumerate}}
\newcommand{\enum}{\end{enumerate}}
\newcommand{\btbl}{\begin{table}}
\newcommand{\etbl}{\end{table}}
\newcommand{\btbu}{\begin{tabular}}
\newcommand{\etbu}{\end{tabular}}
\newcommand{\bcl}{\begin{center}}
\newcommand{\ecl}{\end{center}}
\newcommand{\bbt}{\bibitem}
\newcommand{\beq}{\begin{equation}}
\newcommand{\eeq}{\end{equation}}
\newcommand{\beqr}{\begin{eqnarray}}
\newcommand{\eeqr}{\end{eqnarray}}
\newcommand{\red}{\color{red}}
\newcommand{\blue}{\color{blue}}
\newcommand{\yellow}{\color{yellow}}
\newcommand{\green}{\color{green}}
\newcommand{\purple}{\color{purple}}
\newcommand{\brown}{\color{brown}}
\newcommand{\black}{\color{black}}

\definecolor{boslv}{rgb}{0.0, 0.65, 0.58}
\definecolor{Munsell}{HTML}{00A877}
\newcommand{\psip}{\psi^{'}}
\newcommand{\psipp}{\psi(3686)}

\newcommand{\Br}{\mathcal{B}}
\newcommand{\too}{\rightarrow}
\newcommand{\del}{\color{red}\sout}
\newcommand{\new}{\color{blue}\uwave}
\title{Study of the decay \begin{boldmath}$\mathbf{\psi(3686)}\rightarrow\mathbf{\Sigma^{0}\bar{\Sigma}^{0}\phi}$\end{boldmath}}

\author{
\begin{small}
\begin{center}
M.~Ablikim$^{1}$, M.~N.~Achasov$^{4,c}$, P.~Adlarson$^{76}$, O.~Afedulidis$^{3}$, X.~C.~Ai$^{81}$, R.~Aliberti$^{35}$, A.~Amoroso$^{75A,75C}$, Q.~An$^{72,58,a}$, Y.~Bai$^{57}$, O.~Bakina$^{36}$
% ...
\\
\vspace{0.2cm}
(BESIII Collaboration)\\
\vspace{0.2cm}
\it{
$^{1}$ Institute of High Energy Physics, Beijing 100049, People's Republic of China\\
% ...
$^{3}$ Bochum  Ruhr-University, D-44780 Bochum, Germany\\
$^{4}$ Budker Institute of Nuclear Physics SB RAS (BINP), Novosibirsk 630090, Russia\\
% ...
$^{35}$ Johannes Gutenberg University of Mainz, Johann-Joachim-Becher-Weg 45, D-55099 Mainz, Germany\\
$^{36}$ Joint Institute for Nuclear Research, 141980 Dubna, Moscow region, Russia\\
% ...
$^{57}$ Southeast University, Nanjing 211100, People's Republic of China\\
$^{58}$ State Key Laboratory of Particle Detection and Electronics, Beijing 100049, Hefei 230026, People's Republic of China\\
% ...
$^{72}$ University of Science and Technology of China, Hefei 230026, People's Republic of China\\
% ...
$^{75}$ University of Turin and INFN, (A)University of Turin, I-10125, Turin, Italy; (B)University of Eastern Piedmont, I-15121, Alessandria, Italy; (C)INFN, I-10125, Turin, Italy\\
$^{76}$ Uppsala University, Box 516, SE-75120 Uppsala, Sweden\\
% ...
$^{81}$ Zhengzhou University, Zhengzhou 450001, People's Republic of China\\

\vspace{0.2cm}
$^{a}$ Deceased\\
% ...
$^{c}$ Also at the Novosibirsk State University, Novosibirsk, 630090, Russia\\
% ...
}
\end{center}
\vspace{0.4cm}
\end{small}
}

\date{December 9, 2024}

	\begin{abstract}
		Using $(27.12\pm 0.14)\times 10^{8}$ $\psi(3686)$
                events collected with the BESIII detector operating at
                the BEPCII collider, the decay
                $\psi(3686)\to\Sigma^{0}\bar{\Sigma}^{0}\phi$ is
                observed for the first time with a statistical
                significance of 7.6$\sigma$. Its branching fraction is
                measured to be $(2.64 \pm 0.32_{\textrm{stat}} \pm
                0.12_{\textrm{sys}}) \times 10^{-6}$, where the first
                uncertainty is statistical and the second is
                systematic. In addition, we search for potential
                intermediate states in the $\Sigma^{0}\phi$
                ($\bar{\Sigma}^{0}\phi$) invariant mass distribution
                and a possible threshold enhancement in the
                $\Sigma^{0}\bar{\Sigma}^{0}$ system, but no conclusive
                evidence of is observed.
        \end{abstract} 
\maketitle
	
   \section{Introduction}\label{sec:introduction}\vspace{-0.3cm}

Unlike the well established theory of electromagnetic interactions,
Quantum Chromodynamics (QCD) theory faces challenges in the
non-perturbative region, where the theoretical calculations rely on
approximations and models in various situations. As of today, there is
still no universally accepted and reliable calculation technique of
strong interactions in this region~\cite{QCD_story}.
   
The charmonium states are usually interpreted as bound states
consisting of a charm quark and an anti-charm quark. Experimental
studies of the hadronic decays of charmonium states, which lie between
the perturbative and non-perturbative regimes, are important to test
the current QCD theory and the theoretical
models~\cite{QCD_charmonium1,QCD_charmonium2}. Although the current
QCD-based theoretical framework effectively describes almost all of
the observed hadrons, some of the predicted states have not yet been
found, and exotic states that cannot be conventionally interpreted
as two-quark mesons or three-quark baryons have been observed. The
investigations of the decays of $\psi(3686)\rightarrow B\bar{B}P~(V)$,
where $B$ denotes a baryon, $\bar{B}$ denotes its anti-particle and
$P~(V)$ denotes a pseudoscalar (vector) meson, are essential to search
for intermediate states and excited baryons that have not yet been
observed~\cite{excited,baryons}.
   
In recent years, the decays
$\psi(3686)\rightarrow\Lambda\bar{\Lambda}\pi^{0}$,
$\Lambda\bar{\Lambda}\eta$~\cite{LLeta},
$\Lambda\bar{\Lambda}\eta^{\prime}$~\cite{LLeta_prime},
$\Lambda\bar{\Lambda}\omega$~\cite{LLomega},
$\Sigma^{+}\bar{\Sigma}^{-} \omega$, $\Sigma^{+}\bar{\Sigma}^{-}
\phi$~\cite{SSphi} and $\Sigma^{+}\bar{\Sigma}^{-} \eta$~\cite{SSeta}
have been observed at BESIII. Evidence of the excited state
$\Lambda^{*}$ is found in $\psi(3686)\to\Lambda\bar{\Lambda}\omega$
with a significance of 3.0$\sigma$~\cite{LLomega}, and the excited
state $\Lambda(1670)$ is observed with a significance larger than
5.0$\sigma$ in the $\Lambda\eta$ system in
$\psi(3686)\to\Lambda\bar{\Lambda}\eta$~\cite{LLeta}. Additionally, a
near-threshold enhancement in the $\Lambda\bar{\Lambda} $ invariant
mass spectrum is observed for the first time in
$e^{+}e^{-}\to\Lambda\bar{\Lambda}\phi$~\cite{LLphi}. These exciting
results stimulate us to search for such effects in the similar decay
$\psi(3686)\to\Sigma^{0}\bar{\Sigma}^{0}\phi$.

In this paper, we report the first observation of the decay
$\psi(3686)\to\Sigma^{0}\bar{\Sigma}^{0}\phi$ and measure its
branching fraction using $(27.12\pm 0.14)\times 10^{8}$ $\psi(3686)$
events~\cite{psip_num_21}. In addition, we search for potential
intermediate states in the $\Sigma^{0}\phi~(\bar{\Sigma}^{0}\phi$)
invariant mass spectrum and a possible threshold enhancement in the
$\Sigma^{0}\bar{\Sigma}^{0}$ system.

\section{BESIII Detector and Monte Carlo Simulation}\label{sec:detector}\vspace{-0.3cm}

The BESIII detector~\cite{BES_design} records symmetric $e^+e^-$
collisions provided by the BEPCII storage
ring~\cite{Yu:IPAC2016-TUYA01} in the center-of-mass energy range from
1.84 to 4.95~GeV, with a peak luminosity of $1.1 \times
10^{33}\;\text{cm}^{-2}\text{s}^{-1}$ achieved at $\sqrt{s} =
3.773\;\text{GeV}$.  BESIII has collected large data samples in this
energy region~\cite{Ablikim:2019hff} \cite{EventFilter}. The
cylindrical core of the BESIII detector covers 93\% of the full solid
angle and consists of a helium-based multilayer drift chamber (MDC), a
time-of-flight system (TOF), and a CsI (Tl) electromagnetic
calorimeter (EMC), which are all enclosed in a superconducting
solenoidal magnet providing a 1.0~T magnetic field. The magnet is
supported by an octagonal flux-return yoke with modules of resistive
plate muon counter (MUC) interleaved with steel. The charged-particle
momentum resolution at 1~GeV/$c$ is 0.5\%, and the specific ionization
energy loss d$E/$d$x$ resolution is 6\% for the electrons from Bhabha
scattering at 1~GeV. The EMC measures photon energy with a resolution
of 2.5\% (5\%) at 1~GeV in the barrel (end cap) region. The time
resolution of the TOF plastic scintillator barrel part is 68~ps, while
that of the end cap part was 110~ps. The end cap TOF system was
upgraded in 2015 using multigap resistive plate chamber technology,
providing a time resolution of 60~ps, which benefits $\sim$$83\%$ of
the data used in this analysis~\cite{tof_a,tof_b,tof_c}.
  
Monte Carlo (MC) simulated data samples produced with {\sc
  geant4}-based~\cite{geant4} software, which includes the geometric
description~\cite{detvis} of the BESIII detector and the detector
response, are used to optimize the event selection criteria, determine
the detection efficiencies and study the background components. The
simulation models the beam energy spread and initial-state radiation
in the $e^+e^-$ annihilations using the generator {\sc
  kkmc}~\cite{kkmc_a,kkmc_b}. The inclusive MC sample includes the
production of the $\psi(3686)$ resonance, the initial-state radiation
production of the $J/\psi$, and the continuum processes incorporated
in {\sc kkmc}. All particle decays are modeled by {\sc
  evtgen}~\cite{evtgen_a,evtgen_b} using branching fractions either
taken from the Particle Data Group (PDG)~\cite{pdg2022}, when
available, or otherwise estimated with {\sc
  lundcharm}~\cite{lundcharm_a,lundcharm_b}. Final state radiation
from charged final state particles is included using {\sc
  photos}~\cite{photos}. The signal MC sample, consisting of
$3.0\times 10^{6}$ events, is generated using a phase space model
(PHSP) to determine the detection efficiency.  The data sample
collected at the center-of-mass energy of $\sqrt{s}=3.650$~GeV,
corresponding to an integrated luminosity of 401
pb$^{-1}$~\cite{psip_num_21}, is used to study the continuum
background.

   \section{\label{Sec:Selection}Event Selection and Background Analysis}\vspace{-0.3cm}
   
The decay modes $\Sigma^{0} \to \gamma \Lambda$, $ \Lambda \to p
\pi^{-}$ ($\bar{\Sigma}^{0} \to \gamma\bar{\Lambda}$,
$\bar{\Lambda}\to \bar{p} \pi^{+}$) and $\phi \to K^{+}K^{-}$, which
have large branching fractions and high reconstruction efficiency, are
selected to reconstruct our signal process.  To further increase the
efficiency, we apply a partial reconstruction method, where either
the $\bar{\Sigma}^{0}$ baryon or the $\Sigma^0$ baryon is not
reconstructed but instead identified by the recoiling mass of the
reconstructed particles. The tag method missing the $\bar{\Sigma}^{0}$
is referred to as ``tag A", and the method missing the $\Sigma^0$ is
referred to as ``tag B".  To avoid double counted events, tag B is
used only when tag A is not feasible. For example, if the $\Lambda$
baryon cannot be reconstructed due to a missing proton or $\pi^{-}$
meson, reconstructing the $\Sigma^{0}$ is also not possible, and as a
result, we turn to tag B. The events selected by tag A or tag B are
analysed separately.

Each candidate event is required to contain at least two positive and
two negative charged tracks and at least one candidate
photon. Additionally, the polar angle ($\theta$) of the charged tracks
detected by the MDC must be in the range $\left| \cos \theta\right|
\leq 0.93 $, where $\theta$ is defined with respect to the $z$ axis,
i.e., the symmetry axis of the MDC.

Photon candidates are identified using showers in the EMC. The
deposited energy of each shower must be greater than 25 MeV in the
barrel region ($|\cos\theta|<0.80$) or greater than 50 MeV in the end
cap region ($0.86<|\cos\theta|<0.92$). To suppress electronic noise
and energy depositions not associated with the event, we further
require the EMC cluster time from the reconstructed event start time
to fall within the range $[0,700]$~ns. To veto showers originating
from charged tracks, the opening angle between the extrapolated
trajectory of any charged track and the shower position must exceed 10
degrees.
    
Particle identification~(PID) for charged tracks combines measurements
of d$E$/d$x$ in the MDC and the flight time in the TOF to
calculate the probability $P(h)~(h=p, K, \pi)$ for each particle type
hypothesis. The track with the highest probability of being a proton
is assumed to be a proton, i.e., $P(p)>P(K)$ and $ P(p)>P(\pi)$, while
pions are identified by requiring $P(\pi)>P(K)$ and $
P(\pi)>P(p)$. The remaining tracks that are not identified as
proton or pion are assumed to be kaons originating from the
$\phi$ meson. Each candidate signal event is required to contain
exactly one $K^{+}$ and one $K^{-}$. Additionally, we require
$N_{\pi^{-}}=1$, $N_{p}=1$, $N_{\pi^{+}}< 2$, and $N_{\bar{p}}< 2$ for
candidate events in tag A, and $N_{\pi^{+}}=1$, $N_{\bar{p}}=1$,
$N_{\pi^{-}}<2$, and $N_{p}<2$ for candidate events in tag B, where
$N$ denotes the number of different particles in the event. We limit
the numbers of $p$ and $\pi$ of the recoiling side to suppress the
background.
    
\begin{figure*}[tb]
    \centering
    \mbox{
    \begin{overpic}[width=0.4\textwidth,clip=true]{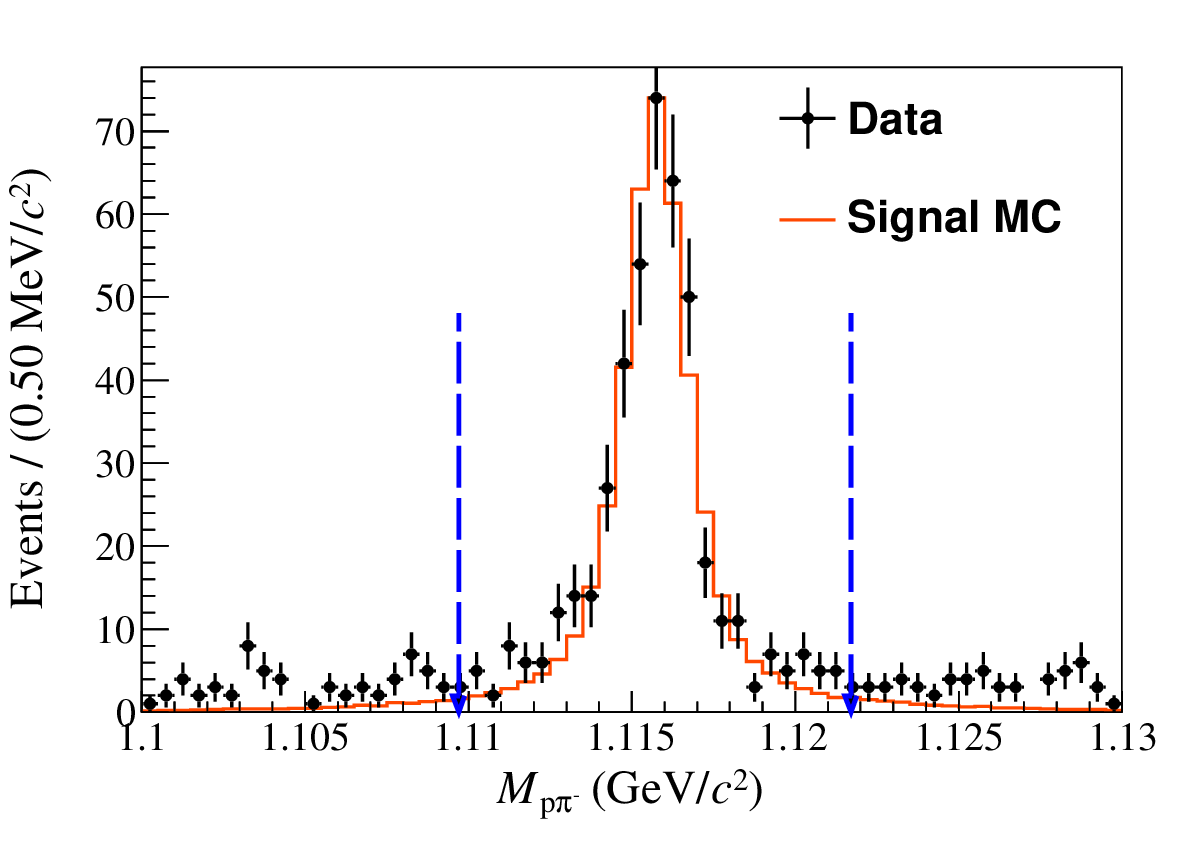}
    \end{overpic}
    \begin{overpic}[width=0.4\textwidth,clip=true]{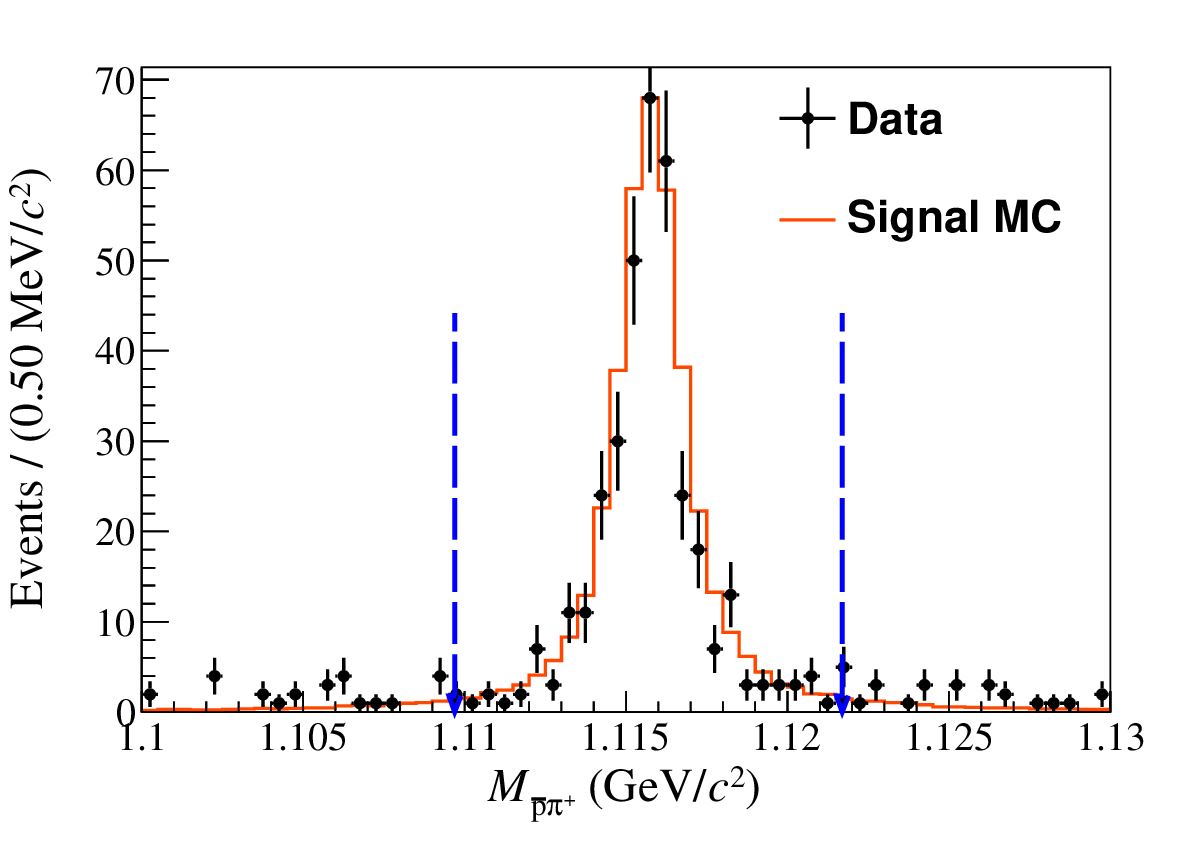}
    \end{overpic}
    }
    \caption{(left) Distribution of $M_{p\pi^{-}}$ for tag A. (right)
      Distribution of $M_{\bar{p}\pi^{+}}$ for tag B. The dots with
      error bars are data, the orange solid lines represent
      the signal MC sample, the blue dashed lines show the
      required range.}
   \label{mlamb}
   \end{figure*}

The charged tracks assumed to be kaons originating from the $\phi$
decay are required to satisfy $\left| V_{z}\right| < 10$ cm and
$\left| V_{xy}\right| < 1$ cm, where $\left|V_{z}\right|$ is the
distance of closest approach to the interaction point (IP) along the
$z$ direction and $V_{xy}$ is the distance in the transverse $x-y$
plane. A vertex fit is performed on the $K^{+}K^{-}$ pair to
reconstruct the $\phi$ meson. A secondary vertex fit is performed on
the $p\pi^{-}$($\bar{p}\pi^{+}$) pair to reconstruct the
$\Lambda$($\bar{\Lambda}$) state. The $\Lambda$($\bar{\Lambda}$) decay
length, defined as the distance of the $\Lambda$($\bar{\Lambda}$)
decay vertex from the IP, is required to be greater than
zero. Additionally, the $\chi^{2}$ value of the secondary vertex fit
is required to be less than 30. The surviving events are required to
fall within
$\left|M_{p\pi^{-}(\bar{p}\pi^{+})}-m_{\Lambda(\bar{\Lambda})}\right|<0.006$~GeV/$c^2$~(four
standard deviations), as shown in Fig.~\ref{mlamb}, where
$M_{p\pi^{-}(\bar{p}\pi^{+})}$ is the invariant mass of the $p\pi$
pair and $m_{\Lambda(\bar{\Lambda})}$ is the
$\Lambda(\bar{\Lambda})$ mass~\cite{pdg2022}.

In the signal channel, the $\Sigma^{0}$ and $\bar{\Sigma}^{0}$ states
always appear in pairs and the proper photon should make the
reconstructed mass of $\gamma\Lambda$ ($\gamma\bar{\Lambda}$) close to
the recoil mass of $\gamma\Lambda\phi$
($\gamma\bar{\Lambda}\phi$). For tag A, the candidate photon is
selected by minimizing the variable $\Delta = (M_{\gamma\Lambda} -
M_{\gamma{\Lambda}\phi}^{\rm{ rec}})^{2}$, where $M_{\gamma\Lambda}$
is the invariant mass of $\gamma\Lambda$ and
$M_{\gamma{\Lambda}\phi}^{\rm rec}$ is the recoiling mass of
$\gamma{\Lambda}\phi$. The photon in tag B is selected similarly by
minimizing $\Delta = (M_{\gamma\bar{\Lambda}} -
M_{\gamma{\bar{\Lambda}}\phi}^{\rm rec})^{2}$.
   
The background components are studied with the $\psi(3686)$ inclusive
MC sample using TopoAna~\cite{zhouxy_topoAna}; the remaining events
can be categorized into eight types depending on their distinctive
peaking characteristics, as summarized in Table~\ref{list_topo}. Here,
``non-$\phi$" indicates that the $\phi$ peak is not observed in the
decay chain of the background, and ``non-$\Sigma^{0}$" and
``non-$\bar{\Sigma}^{0}$'' have similar meanings. Type 6 and type 7
events are the dominant background sources, while the contributions
from types 1-5 are relatively negligible. Specifically, the dominant
background channels of type 6 events are the decays $\chi_{cJ}\,
(J=0,\,1,\, 2)\to \Lambda \bar{\Lambda} \phi$, while for type 7 events
they are the decays $\chi_{cJ}\to \bar{\Omega}^{+} \Omega^{-}$.

 \begin{table}[tb]
        \centering        
        \caption{Numbers of surviving events in the $\psi(3686)$ inclusive MC sample after selections.}
        \label{list_topo}
\begin{tabular}{cccccc}
\hline
\hline

      \multicolumn{4}{c}{Event type }  &Tag A        & Tag B      \\  \hline
       type 0  \quad   &  $\phi$ \quad & $\Sigma^{0}$ \quad &  $\bar{\Sigma}^{0}$ \quad & 452            &  430       \\
       type 1  \quad     &    non-$\phi$ \quad & $\Sigma^{0}$ \quad & $\bar{\Sigma}^{0}$         \quad &1 &        1\\
        type 2  \quad    &        $\phi$ \quad & $\Sigma^{0}$ \quad & non-$\bar{\Sigma}^{0}$       \quad &44      &     40    \\
        type 3  \quad    &  $\phi$ \quad & non-$\Sigma^{0}$ \quad & $\bar{\Sigma}^{0}$          \quad &46     &      46   \\
        type 4  \quad    & non-$\phi$ \quad & $\Sigma^{0}$ \quad &non-$\bar{\Sigma}^{0}$            \quad &18      &     1    \\
        type 5  \quad    & non-$\phi$ \quad & non-$\Sigma^{0}$ \quad & $\bar{\Sigma}^{0}$            \quad &   4    &      10   \\
        type 6  \quad    & $\phi$ \quad & non-$\Sigma^{0}$ \quad & non-$\bar{\Sigma}^{0}$            \quad &6090     &   4866      \\
        type 7  \quad    & non-$\phi$ \quad & non-$\Sigma^{0}$ \quad & non-$\bar{\Sigma}^{0}$           \quad &335      &     246    \\
        
\hline
\hline

\end{tabular}
\end{table}

The yields of the continuum background events are estimated using the
data sample collected at $\sqrt{s}=3.650$~GeV. Only a few events survive and
no peaking contribution is observed, indicating that the continuum
background is negligible.

   \section{\label{Sec:BR_determined}Branching Fraction Measurement}\vspace{-0.3cm}
   
A topological analysis has shown the presence of non-negligible
peaking background in the $\Sigma^{0}$, $\bar{\Sigma}^{0}$ and $\phi$
mass distributions. Below, the invariant mass of
$K^{+}K^{-}$ is referred to as $M_{\phi}$, and
$M_{\gamma\Lambda}$, $M^{\rm rec}_{\gamma\Lambda\phi}$ and so on have
similar meanings. In
addition, a correlation is observed between $M_{\Sigma^{0}}$ and
$M_{\bar{\Sigma}^{0}}$, due to the photon selection. We perform an
extended unbinned maximum likelihood simultaneous fit on the
$M_{\phi}$ distribution and on the $M_{\Sigma^{0}/\bar{\Sigma}^{0}}$
two-dimensional distribution of the accepted candidate events in the
data to determine the number of signal events. Table~\ref{list_model}
lists the probability distribution function (PDF) models used in the
fit. Here, type 0 denotes the signal, ``mismatch" denotes the
background from incorrect combinations, type 2, type 3 and type 6
events are identical to those in Table~I, type 8 incorporates type 1,
type 4 and type 5 events of Table~I, denoting all the non-$\phi$
backgrounds. The $M_{\phi}$ and $M_{\Sigma^{0}/\bar{\Sigma}^{0}}$
signal shapes are derived from the signal MC sample with the
RooKeysPdf and RooNDKeysPdf~\cite{keyspdf} functions,
respectively. The non-$\phi$ background shape is described by a
reverse ARGUS~\cite{ARGUS} function, defined as
\begin{equation}
\mathrm{ARGUS}\left(m;m_0,\xi, p\right)= \begin{cases}0 & \left(m<m_0\right) \\ m \cdot \nu^p \cdot e^{\xi \cdot \nu} & \left( m \geq m_0\right) \end{cases},
\end{equation}
 where $m$ is the $K^{+}K^{-}$ invariant mass, $m_0$ is the mass
 threshold, $\nu={(\frac{m}{m_0})}^2-1$, and $p$ and $\xi$ are free
 parameters.  For type 2, type 3 and type 6 events, the
 $M_{\Sigma^{0}/\bar{\Sigma}^{0}}$ distributions are derived from the
 inclusive MC sample with the RooNDKeysPdf function. For type 8 
 events, the $M_{\Sigma^{0}/\bar{\Sigma}^{0}}$ distributions are
 derived from the $\phi$ sideband events of the data sample with the
 RooNDKeysPdf function.
 \begin{table}[tb]
        \centering        
        \caption{The PDF models used in the simultaneous fit. In the Table, ``signal MC" and ``inclusive MC" correspond to PDF models derived from the MC simulation shapes, while ``$\phi$ sideband'' PDF is derived from the data.} 
\label{list_model}
\begin{tabular}{cccccc}
\hline
\hline

      \multicolumn{4}{c}{Event type } & $x$ ($M_{\phi}$)    &  $y/z$ ($M_{\Sigma^{0}/\bar{\Sigma}^{0}}$)     \\  \hline
       type 0   \quad    &  $\phi$ \quad & $\Sigma^{0}$ \quad & $\bar{\Sigma}^{0}$ \quad &  signal MC        & signal MC      \\
       mismatch    \quad    &   \multicolumn{3}{c}{wrong combinations} \quad   & signal MC &       signal MC\\
       type 2    \quad  &     $\phi$ \quad & $\Sigma^{0}$ \quad &  non-$\bar{\Sigma}^{0}$ \quad            & signal MC      &  inclusive MC   \\
       type 3    \quad  & $\phi$ \quad & non-$\Sigma^{0}$ \quad & $\bar{\Sigma}^{0}$ \quad           & signal MC  &     inclusive MC   \\
       type 6    \quad  & $\phi$ \quad & non-$\Sigma^{0}$ \quad & non-$\bar{\Sigma}^{0}$   \quad         &signal MC  & inclusive MC   \\
       type 8    \quad  & \multicolumn{3}{c}{non-$\phi$} \quad         &ARGUS     &  $\phi$ sideband \\
        
\hline
\hline

\end{tabular}
\end{table}

\begin{figure*}[tb]
   \centering
   \mbox{
    \begin{overpic}[width=0.3\textwidth,clip=true]{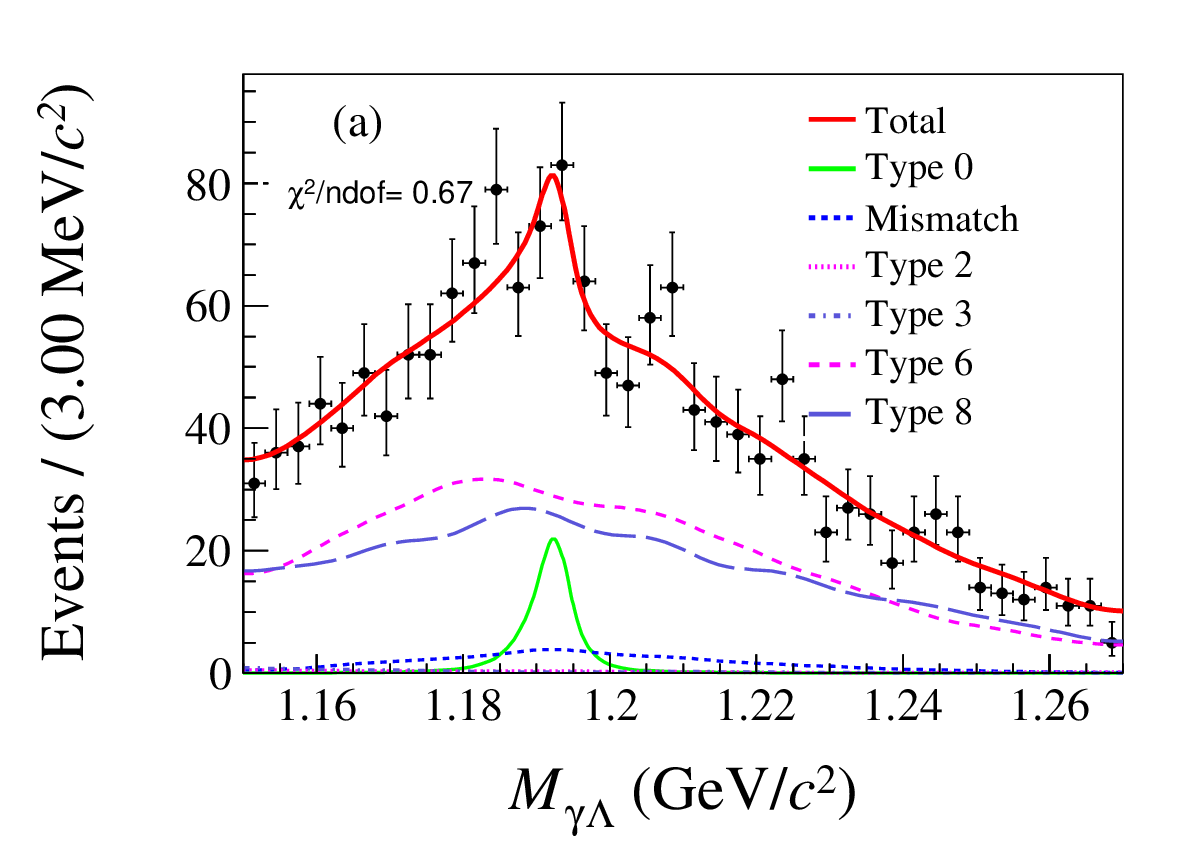}
    \end{overpic}
    \begin{overpic}[width=0.3\textwidth,clip=true]{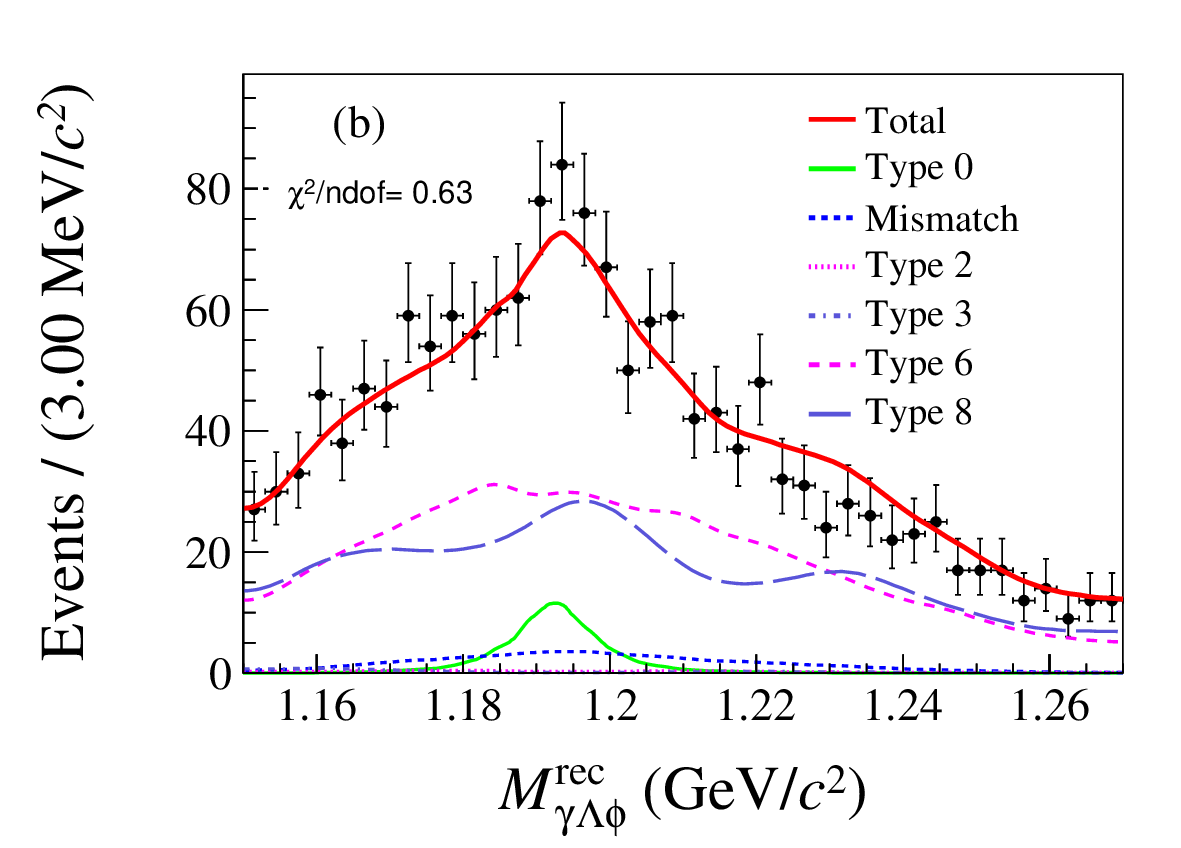}
    \end{overpic}
    \begin{overpic}[width=0.3\textwidth,clip=true]{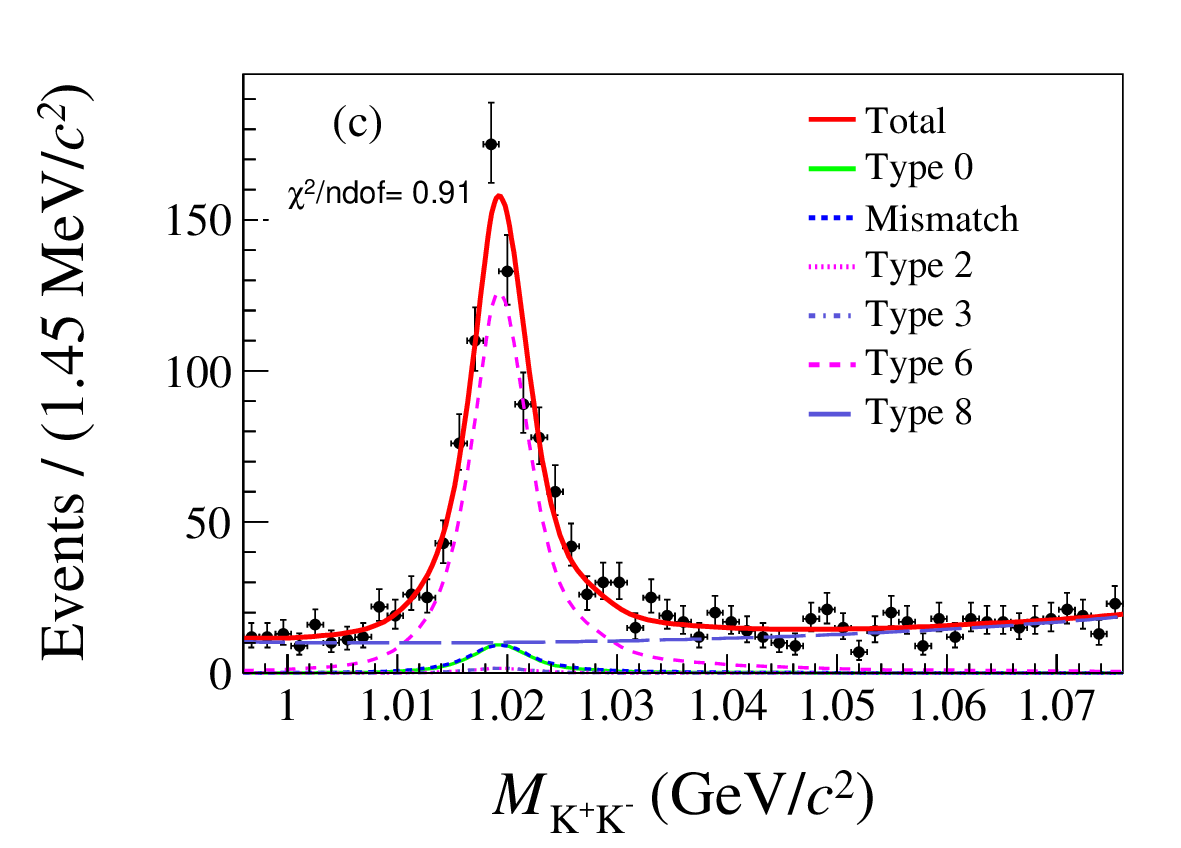}
    \end{overpic}
   }

    \mbox{
    \begin{overpic}[width=0.3\textwidth,clip=true]{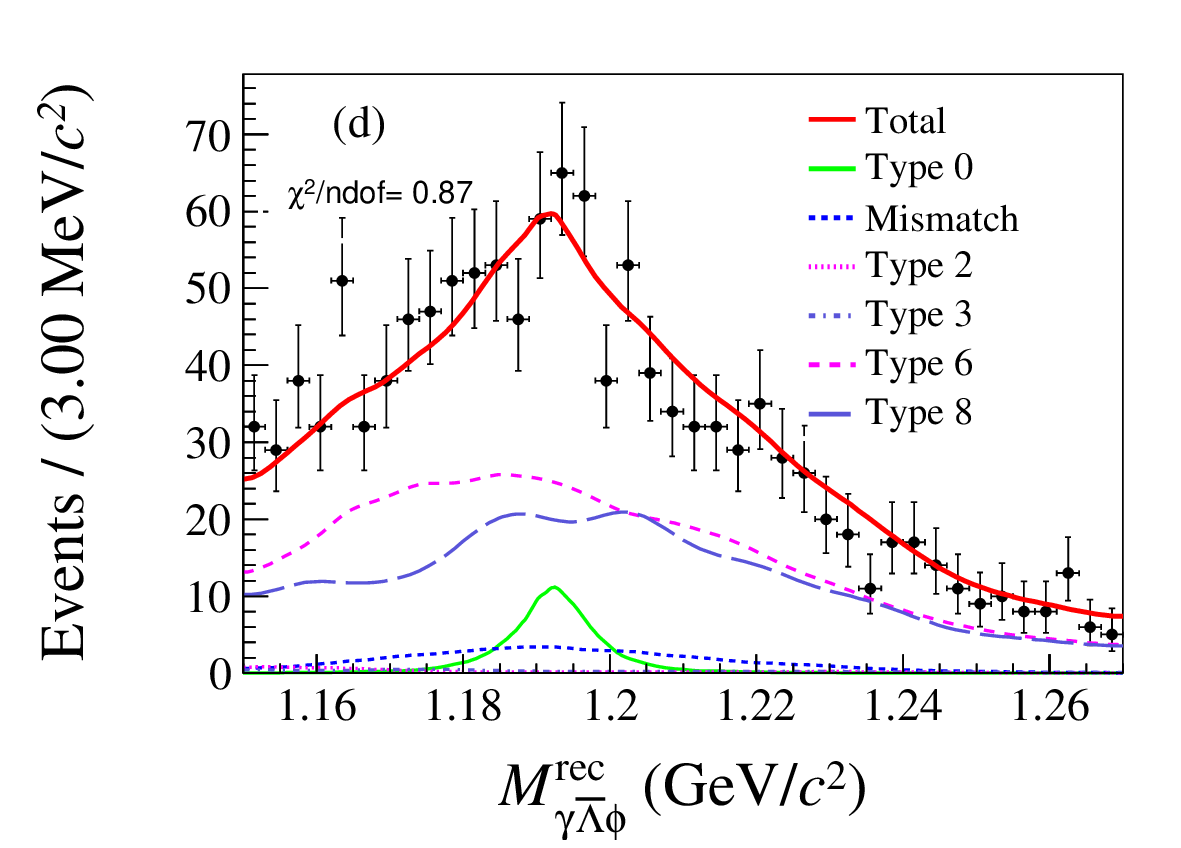}
    \end{overpic}
    \begin{overpic}[width=0.3\textwidth,clip=true]{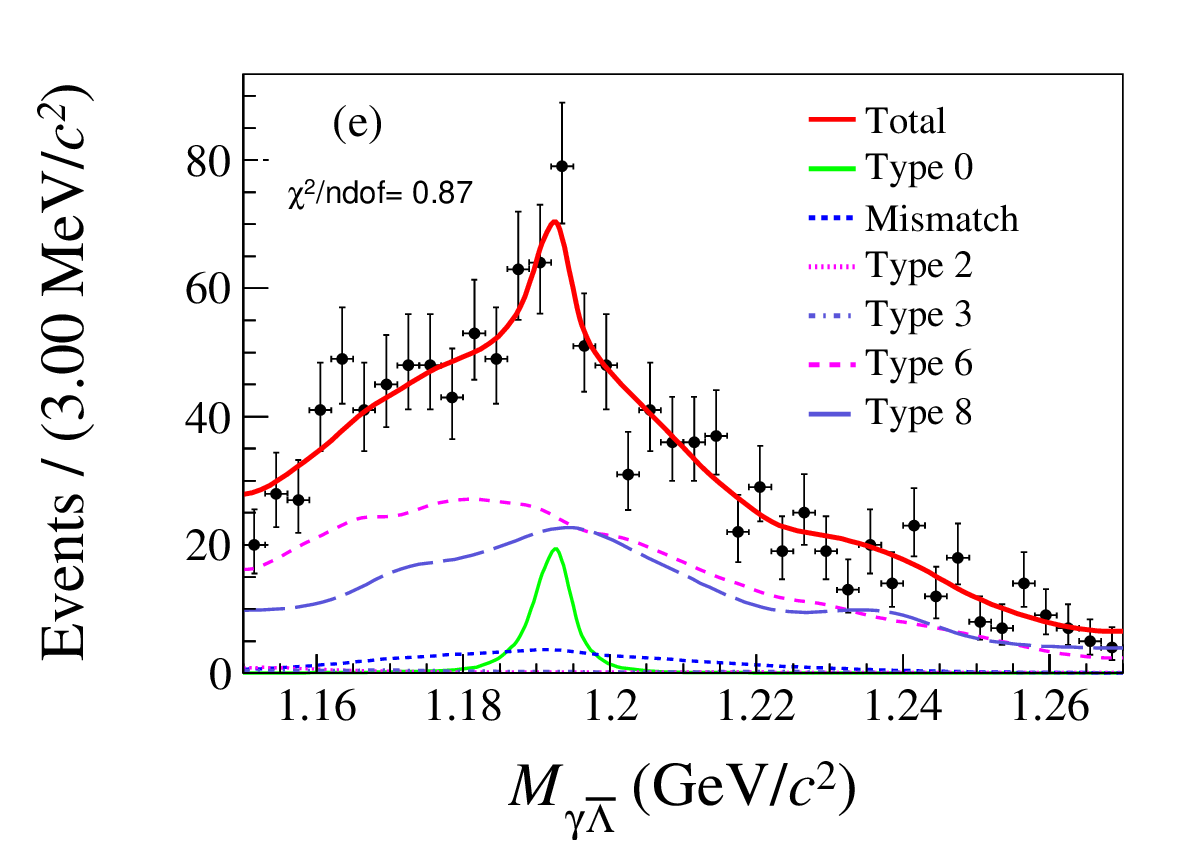}
    \end{overpic}
    \begin{overpic}[width=0.3\textwidth,clip=true]{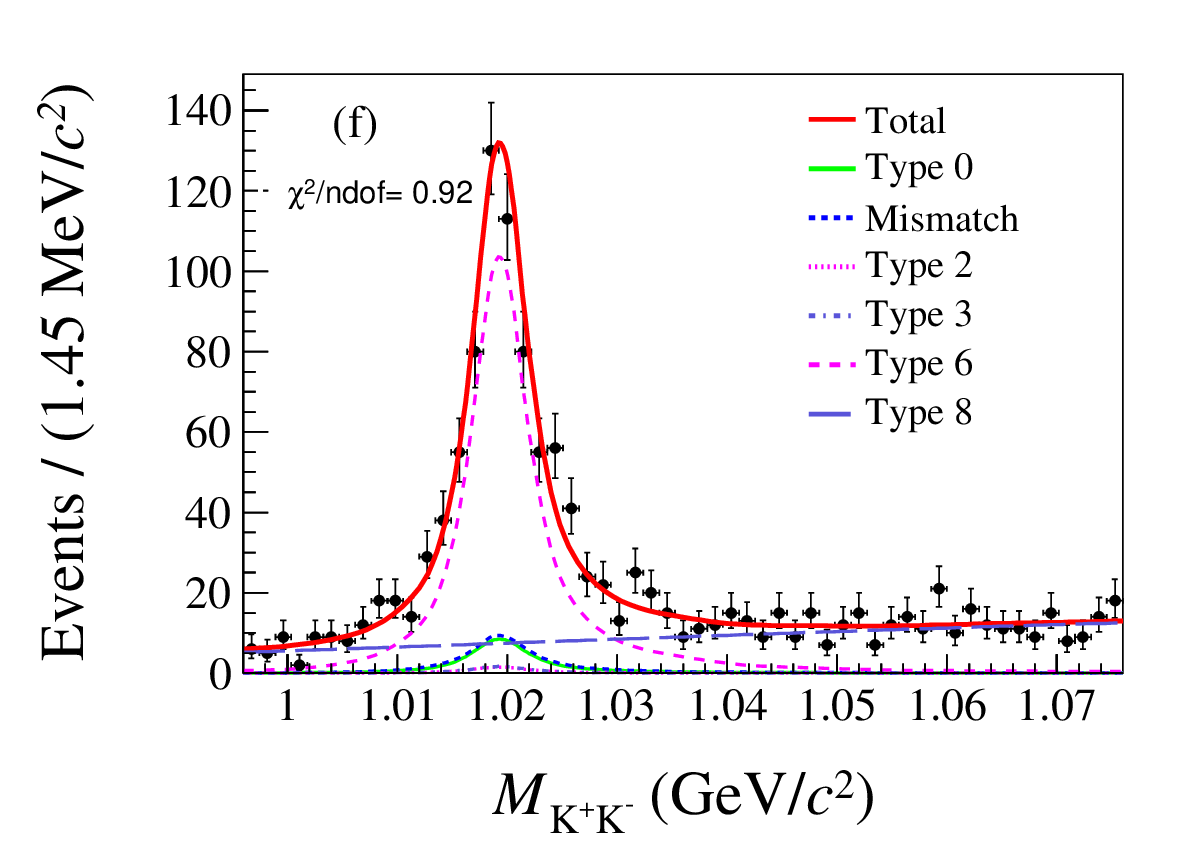}
    \end{overpic}
   }
   \caption{(top) Fit results of (a) $M_{\gamma\Lambda}$ and the
     projections of (b) $M^{\rm rec}_{\gamma{\Lambda}\phi}$ and (c)
     $M_{K^{+}K^{-}}$ for tag A. (bottom) Fit results of (d) $M^{\rm
       rec}_{\gamma\bar{\Lambda}\phi}$ and the projections of (e)
     $M_{\gamma\bar{\Lambda}}$ and (f) $M_{K^{+}K^{-}}$ for tag B. The
     dots with error bars represent the data, the red solid lines
     represent the total fit, the green solid lines represent the
     signal events, the dark blue short dashed lines represent the
     mismatched events, the pink dotted lines and the dark blue
     dash-dotted lines represent the type 2 and type 3 events,
     respectively, the pink dashed lines represent the type 6 events,
     and the dark blue long dashed lines represent the type 8 events.}
   \label{sim_fit}
   \end{figure*}

In the simultaneous fit, the numbers of type 2 and type 3 events are
fixed according to their fractions derived from the inclusive MC
samples, and normalized to the data sample. Additionally, the ratio
between the numbers of mismatched events and correct events is fixed
to its value in the signal MC sample. The fit result is shown in
Fig.~\ref{sim_fit}.
   
From this fit, the number of signal events is determined to be 57.3
$\pm$ 9.7 for tag A and 50.8 $\pm$ 9.1 for tag B.  The statistical
significances of tag A and tag B are estimated from the differences in
the log-likelihood values with and without the signal function, and
are 6.2$\sigma$ and 6.3$\sigma$, respectively.  The significance of
the combined signal from tag A and tag B is $7.6\sigma$.

The branching fraction of
$\psi(3686)\to\Sigma^{0}\bar{\Sigma}^{0}\phi$ is calculated as
   \begin{equation}
   \Br(\psi(3686)\to\Sigma^{0}\bar{\Sigma}^{0}\phi) = 
    {{{{N}_{\mathrm{sig}}^{\mathrm{obs}} \over {N_{\psi \left( {3686} \right)} 
   \cdot \varepsilon_{\mathrm {sig}} \cdot \Br_{\phi} } }}},
   \end{equation}
where ${N_{\mathrm{sig}}^{\mathrm{obs}}}$ is the number of signal
events, $\Br_{\phi}$ is the branching fraction of $\phi \to K^{+}
K^{-}$~\cite{pdg2022},
$\varepsilon_{\mathrm{sig}}$ is the signal efficiency determined by
the MC simulation and the number of $\psi(3686)$ events in data
is ${N_{\psi \left( {3686} \right)}=(27.12\pm
  0.14)\times10^{8}}$~\cite{psip_num_0912}. The branching fraction of
$\Sigma^{0}\to \gamma\Lambda$ is 100\% according to the PDG, thus it
is not included in Eq.~(2). The branching fractions of $\Lambda\to p
\pi^{-}$ and $\bar{\Lambda}\to \bar{p}\pi^{+}$ are not in the formula
either, as they have been incorporated into the signal efficiency. The
weighted statistical uncertainty ($\sigma_{\textrm{stat}}$) and the
average value ($\mu$) of tag A and tag B is calculated as
    \begin{equation}
    \begin{aligned}
   & \frac{1}{\sigma^2_{\textrm{stat}}}=\frac{1}{\sigma_{A}^2}+\frac{1}{\sigma_{B}^2} \quad \text{and}\quad    \mu=\sigma^{2}_{\textrm{stat}}\left(\frac{\mu_{A}}{\sigma_{A}^2}+\frac{\mu_{B}}{\sigma_{B}^2}\right), \\
  \end{aligned}
  \end{equation}
   where the subscripts ``$A$'' and ``$B$'' represent tag A and tag B accordingly. Table~\ref{br_result} lists the numerical results.
    \begin{table}[tb]
        \centering        
        \caption{Signal yield in the data, statistical significance, detection efficiency and branching fraction.}
   \label{br_result}
\begin{tabular}{lcccc}
\hline
\hline
& Signal yield&~Significance~& ~$\varepsilon$ (\%)~ & $\Br~(\times10^{-6})$ \\ 
\hline
Tag A                        & 57.3 $\pm$ 9.7 &6.2$\sigma$        & 1.69                  & 2.54 ± 0.43    \\ 

Tag B           & 50.8 $\pm$ 9.1 & 6.3$\sigma$   & 1.38                  & 2.76 ± 0.50    \\ 

Average~  & -     &-                & -                     & 2.63 ± 0.32   \\ 

\hline
\hline
\end{tabular}
\end{table}

   \section{\label{Sec:inter_states}Study of Intermediate States}\vspace{-0.3cm}

\begin{figure*}[tb]
   \centering
   \begin{minipage}[t]{0.32\linewidth}
   \includegraphics[width=1\textwidth]{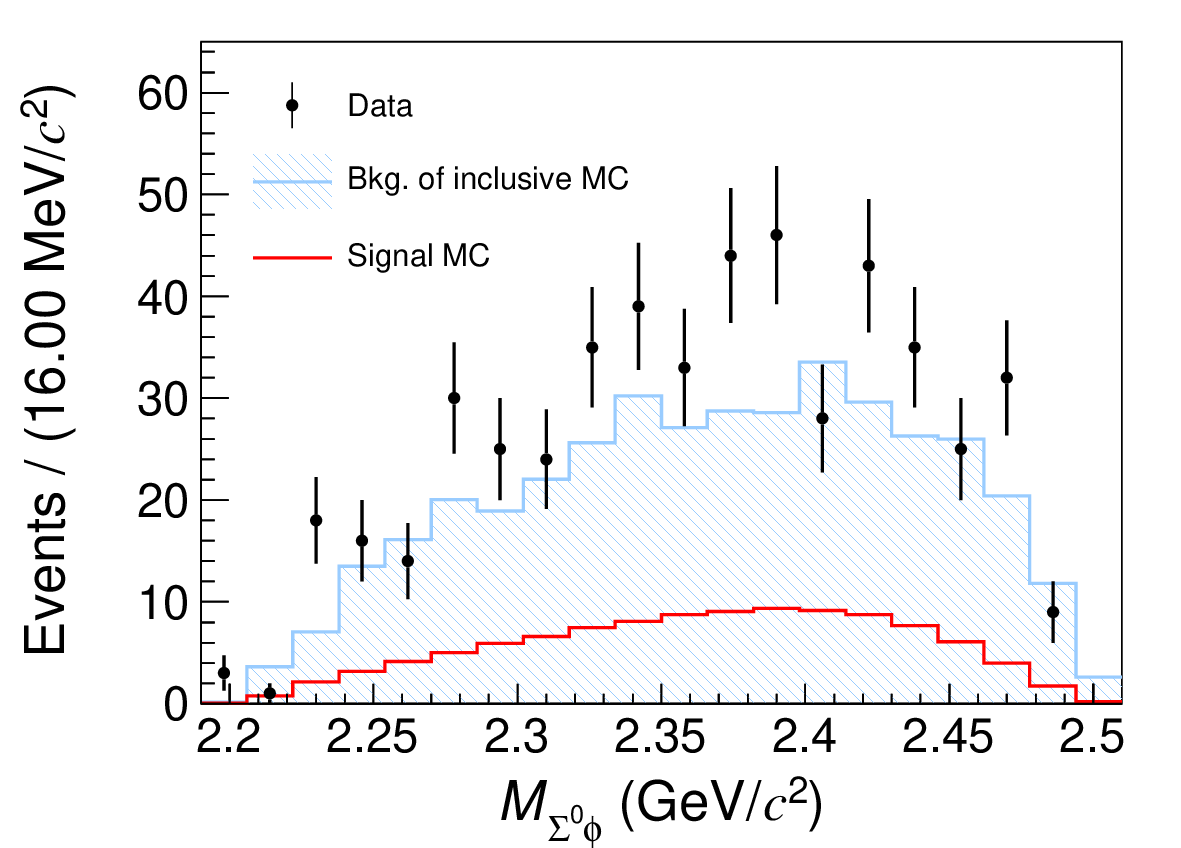}
  \end{minipage}
  \begin{minipage}[t]{0.32\linewidth}
  \includegraphics[width=1\textwidth]{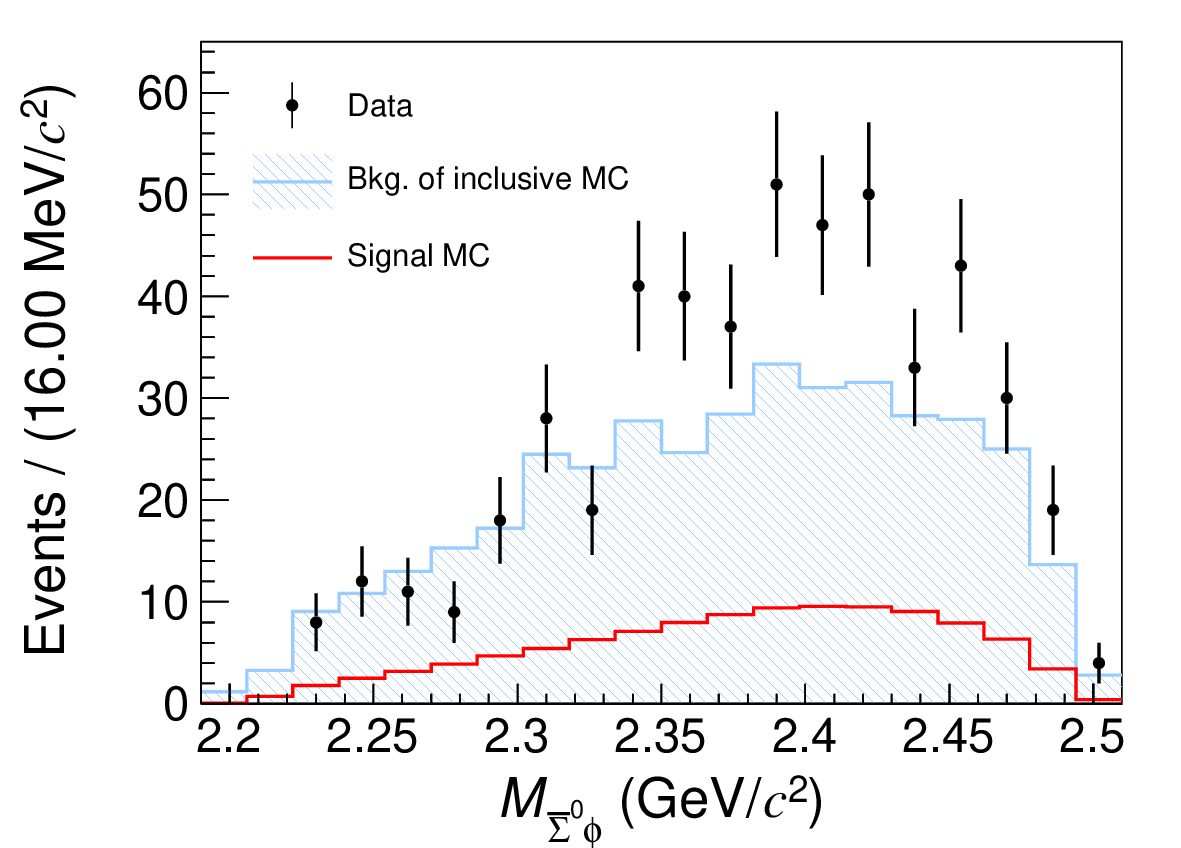}
  \end{minipage}
  \begin{minipage}[t]{0.32\linewidth}
  \includegraphics[width=1\textwidth]{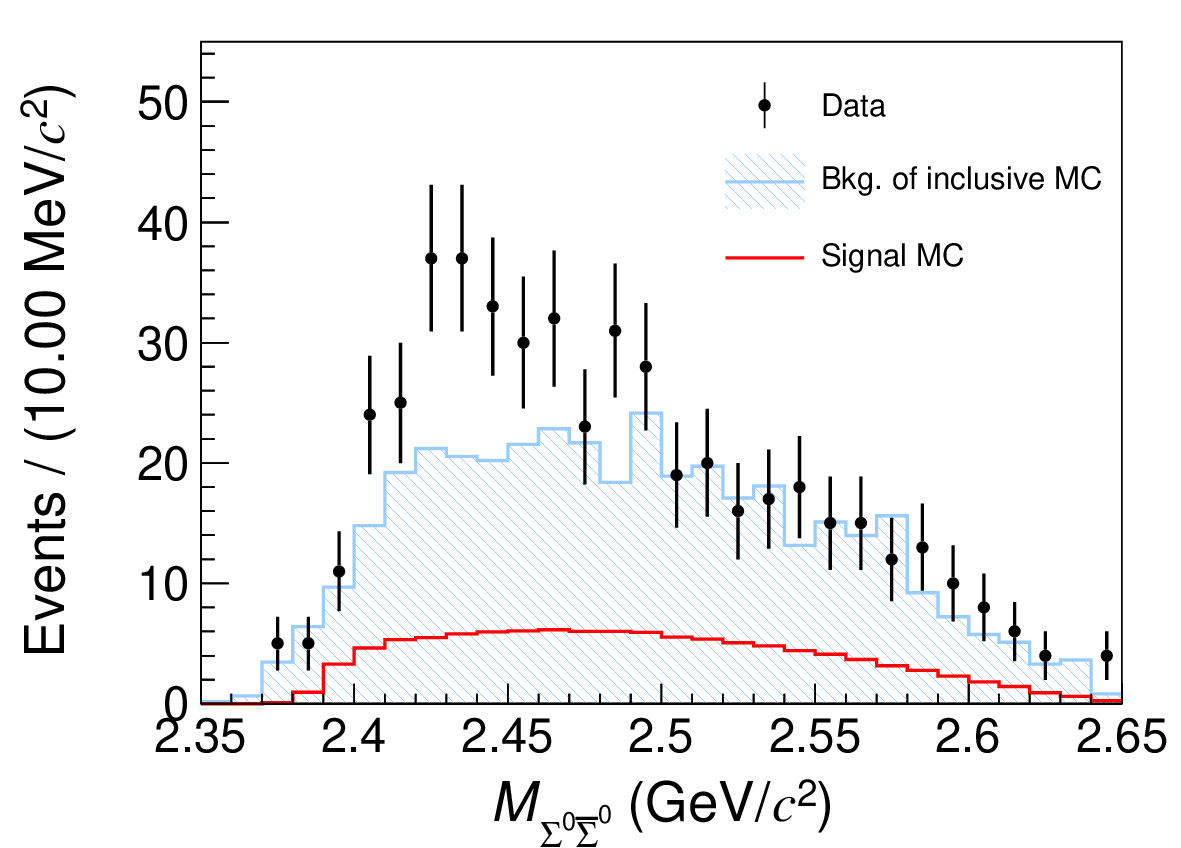}
  \end{minipage}
  
  \begin{minipage}[t]{0.28\linewidth}
  \includegraphics[width=1\textwidth]{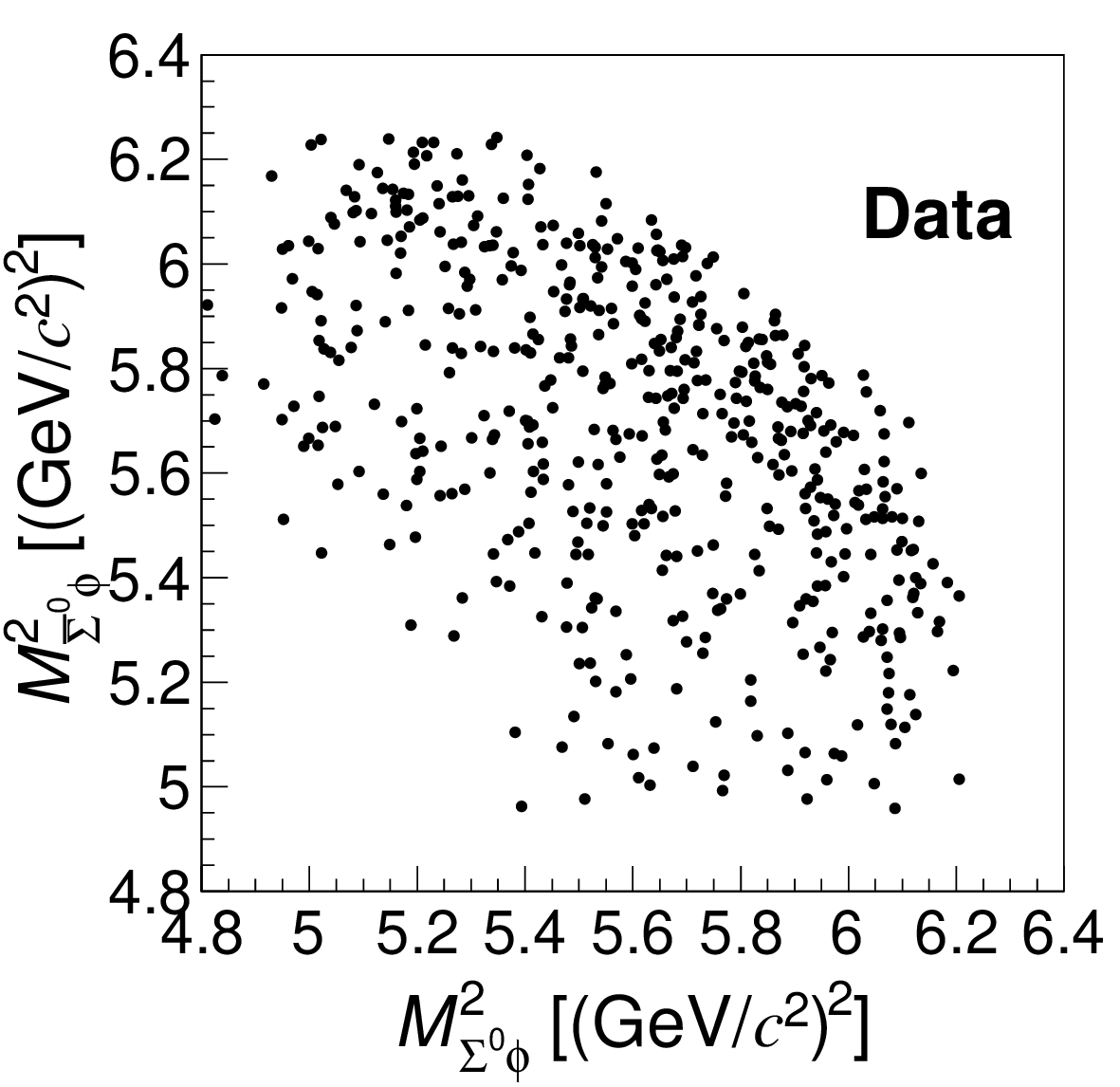}
  \end{minipage}
  \begin{minipage}[t]{0.28\linewidth}
  \includegraphics[width=1\textwidth]{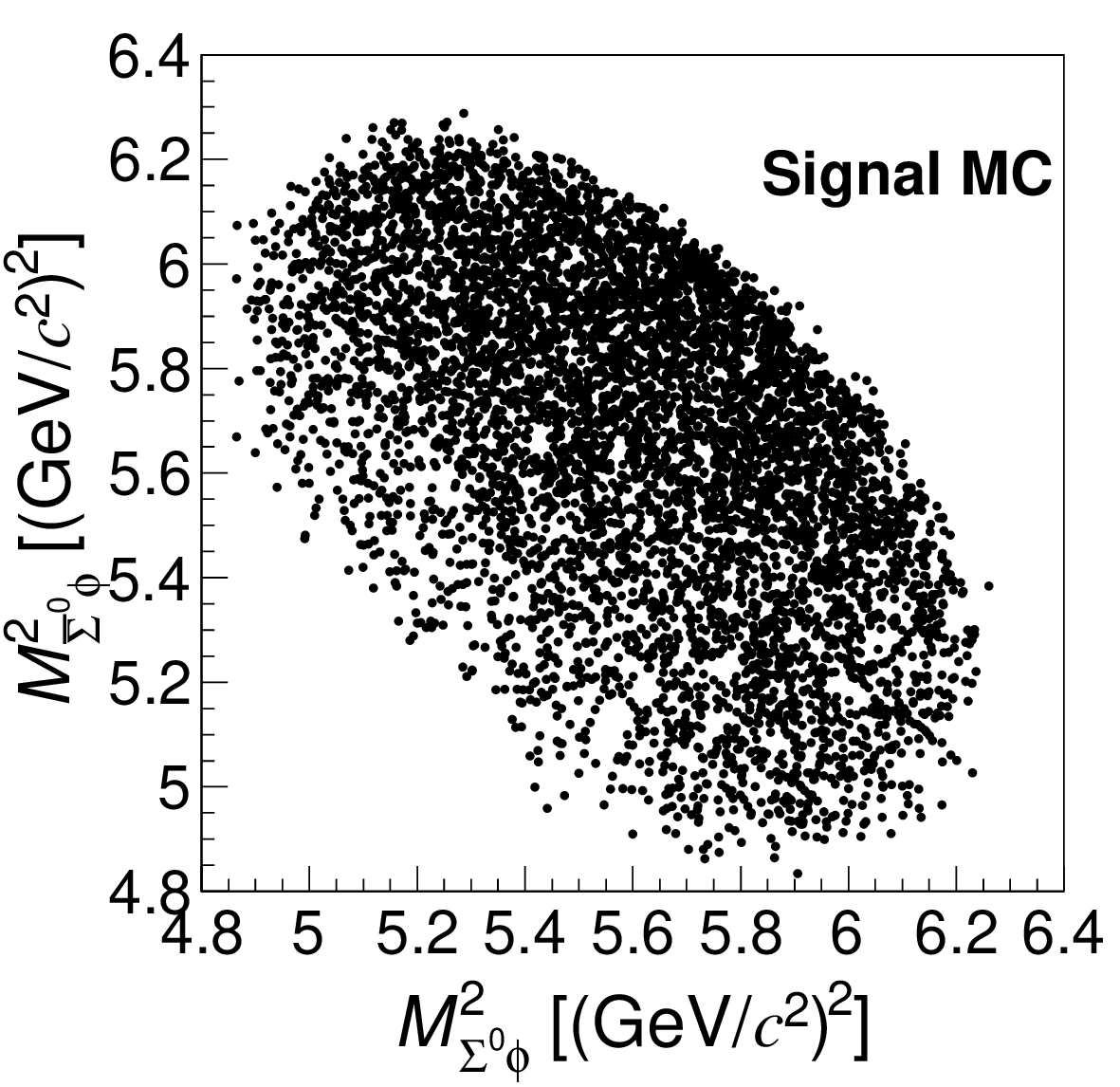}
  \end{minipage}

   \caption{(top) Distributions of $M_{\Sigma^{0}\phi}$,
     $M_{\bar{\Sigma}^{0}\phi}$ and $M_{\Sigma^{0}\bar{\Sigma}^{0}}$
     in the data and MC samples. The black dots with error bars are
     data, the red curves denote the PHSP MC simulated shape and
     the shadowed blue areas denote the backgrounds obtained from the
     inclusive MC sample. The signal and background are normalized
     according to the numbers of events in the data. (bottom) Dalitz
     plots of ${M^{2}_{\Sigma^{0}\phi}}$ versus
     ${M^{2}_{\bar{\Sigma}^{0}\phi}}$ in the data and signal MC
     samples.}
   \label{inter_plots}
   \end{figure*}

In order to investigate potential intermediate states in $\psi(3686)
\rightarrow \Sigma^{0}\bar{\Sigma}^{0} \phi$, candidate events of tag
A and tag B are required to fall within the $\Sigma^0$,
$\bar{\Sigma}^0$, and $\phi$ signal regions, as shown in
Table~\ref{mass_windows}. These signal regions are within three
standard deviations around individual fitted masses. Tag A and Tag B
are combined to increase the statistics. The distributions of
$M_{\Sigma^0\phi}$, $M_{\bar{\Sigma}^0\phi}$ and
$M_{\Sigma^0\bar{\Sigma}^0}$, as well as the Dalitz plots of
${M^{2}_{(\Sigma^{0}\phi)}}$ versus
${M^{2}_{(\bar{\Sigma}^{0}\phi)}}$, are shown in
Fig.~\ref{inter_plots}. In the $M_{\Sigma^{0}\bar{\Sigma}^{0}}$
spectrum, a difference between the PHSP MC and the data is observed
around 2.43~GeV. However, due to the limited statistics, we are unable
to draw a definitive conclusion. Additional data and further research
are required.
    
    \begin{table}[tb]
        \centering        
        \caption{Signal regions of tag A and tag B.}
        \label{mass_windows}
        \begin{tabular}{ccc}
        \hline
        \hline
             &  &Signal region (GeV/$c^2$) \\
             \hline
           \multirow{3}{*}{Tag A}  &  $M_{\gamma\Lambda}$  & [1.176, 1.207]\\
            &$M_{\gamma{{\Lambda}}\phi}^{\rm rec}$  & [1.174, 1.210] \\
             & $M_{K^{+}K^{-}}$& [1.006, 1.032] \\
               \hline
           \multirow{3}{*}{Tag B}&$M_{\gamma{\bar{\Lambda}}\phi}^{\rm rec}$   & [1.175, 1.209]  \\
            & $M_{\gamma\bar{\Lambda}}$&[1.177, 1.207] \\
            & $M_{K^{+}K^{-}}$ & [1.006, 1.032] \\
        \hline
        \hline
        \end{tabular}
    \end{table}   
   
   \section{Systematic Uncertainties}\label{sec:sysU}\vspace{-0.3cm}
   
   The sources of systematic uncertainties on the measured branching fraction are described below. 

\begin{itemize}

\item[(i)]{{\bf Kaon tracking}}

The uncertainty of kaon tracking is assigned as 1.0\% per kaon from the study of the control sample $J/\psi \rightarrow K^0_S K^\pm \pi^\mp$~\cite{Ablikim:2011kv}. 

\item[(ii)]{\bf Photon detection}

The uncertainty of photon reconstruction is assigned as 0.5\% per photon based on the study of the control sample $e^{+}e^{-}\to\gamma\mu^{+}\mu^{-}$~\cite{photon_rec}.

\item[(iii)]{ $\mathbf{\Lambda(\bar{\Lambda})}$ \textbf{reconstruction}}

The uncertainty of $\Lambda(\bar{\Lambda})$ reconstruction includes
the uncertainty of tracking and PID for $p$ and $\pi$, along with the
decay length and the $\chi^{2}$ value requirements. The efficiency of
$\Lambda(\bar{\Lambda})$ reconstruction depends on its polar angle and
momentum distributions, and the distributions of the signal MC sample
differ slightly from those of the data sample. We use the control
sample $J/\psi \rightarrow p K^{-} \bar{\Lambda} + c.\,c
$~\cite{rec_lam(lamb)} to correct the efficiency and estimate the
uncertainty, which is determined to be 1.0\% (0.8\%) for tag A (B).

\item[(iv)]{$\mathbf{\Lambda(\bar{\Lambda})}$ \textbf{mass window}}

We fit the data sample using the simulated shape convolved with a
Gaussian resolution function. After smearing this Gaussian resolution
function to the signal MC sample, the change in the efficiency is
estimated as the related uncertainty, and it is 0.2\% (0.2\%) for tag A
(B).

\item[(v)]{\bf Wrong combination ratio}

The wrong combination ratio of the signal MC sample may vary  from that of the data sample. We adjust the matching angle in the MC truth association by $\pm 1^{\circ}$~\cite{ratio} and take the larger change of the re-measured branching fractions as the systematic uncertainty, which is 0.8\% (1.4\%) for tag A (B).

\item[(vi)]{\bf MC sample size}

The relative uncertainty of the signal MC sample size is estimated as
$\sqrt{\frac{1-\varepsilon}{N \cdot \varepsilon}}
$, where $\varepsilon$ is the detection efficiency and $N$ is the total number of the generated signal events. It is 0.4\% (0.5\%) for tag A (B).

\item[(vii)]{\bf Fit procedure}

  \begin{itemize}
  \item[\textbullet] {\bf Fit range}:
we simultaneously vary the fit range of $M_{\Sigma^{0}}$, $ M_{\bar{\Sigma}^{0}}$ and $M_{\phi}$ by $\pm$ 10~MeV/$c^2$, except for the lower limit of $M_{\phi}$, as there is a threshold. The larger difference in the re-measured branching fraction is assigned as the uncertainty, which is 4.5\% (2.6\%) for tag A (B).
   
  \item[\textbullet]  {\bf Fixed numbers of background events}: assuming that the number of type 2 events follows a Poisson distribution, we adjust it by $\pm \sqrt{N_2}$, where $N_2$ denotes the number of type 2 events that is fixed in the fit. Alternative fits are performed with modified $N_2$ and the larger difference in the branching fraction is assigned as the uncertainty. It is 0.6\% (0.4\%) for tag A (B). The uncertainty associated with the number of type 3 events is estimated similarly, which is 0.6\% (0.2\%) for tag A (B).

  \item[\textbullet]  {\bf Signal shape}:
the description of the $\phi$ shape is changed from the RooKeysPdf
function to the RooHistPdf function. At the same time, the overall
scale factor of the bandwidth of the RooNDKeysPdf function is varied
from 1.0 to 1.3, representing the uncertainty of the $M_{\Sigma^{0}/\bar{\Sigma}^{0}}$ shape. The uncertainty associated with the signal shape is assigned as 1.7\% (0.6\%) for tag A (B).

  \item[\textbullet] {\bf Background shape}: the uncertainties of the
    shapes of the type 2 and type 3 events are negligible. For the
    shape of type 8 events, the smooth shape of $M_{K^{+}K^{-}}$ is
    replaced by a second order Chebyshev polynomial function, and the
    $M_{\Sigma^{0}/\bar{\Sigma}^{0}}$ two-dimensional distribution is
    changed by moving the $\phi$ sideband range by 10~MeV/$c^{2}$
    towards the higher mass region. As for the shapes of the type 6
    and mismatched events, their $\phi$ shapes are changed to the
    RooHistPdf function instead of the RooKeysPdf function, and the
    distributions of $M_{\Sigma^{0}/\bar{\Sigma}^{0}}$ are varied by
    changing the bandwidth scale factor of the RooNDKeysPdf function
    from 1.0 to 1.3. The difference in the branching fraction after
    modifying the background shape is considered as the uncertainty,
    and it is 1.4\% (3.5\%) for tag A (B).
 \end{itemize}
 
\item[(viii)]{\bf Quoted branching fractions}

   The uncertainties of $\Br\left(\Lambda\to p \pi^{-}/\bar{\Lambda}\to \bar{p} \pi^{+}\right)$ and $\Br(\phi \to K^{+} K^{-})$ are taken from the PDG~\cite{pdg2022}, which are 0.8\% and 1.0\%, respectively.

\item[(ix)]{\bf Number of $\psi(3686)$ events}

 The uncertainty of the total number of $\psi(3686)$ events in the data sample is 0.5\%~\cite{psip_num_0912,psip_num_21}.
\end{itemize}

The correlation of the uncertainties between tag A and tag B are
considered. We divide the uncertainties into uncorrelated
individual uncertainties and correlated common uncertainties, which
include kaon tracking, photon detection, quoted branching fractions,
number of $\psi(3686)$ events, and $\Lambda(\bar{\Lambda})$
reconstruction. For a conservative estimation, we assume full
correlation for the correlated uncertainties. All the systematic
uncertainty sources and their values are summarized in
Table~\ref{uncertainty}.

\begin {table}[tb]
\renewcommand\arraystretch{1.2}
{\caption {Systematic uncertainties of the branching fraction measurement.}
\label{uncertainty}}
\begin{tabular}{lcc}
\hline
\hline

        Source  &         Tag A (\%)           & Tag B (\%)      \\  \hline
        Kaon tracking& 2.0            &      2.0   \\
        Photon detection     &  0.5 &    0.5     \\
        $\Lambda$ reconstruction & 1.0      &     -    \\
        $\bar{\Lambda}$ reconstruction & -     &   0.8      \\
        $\Lambda$ mass window &   0.2     &     -   \\
        $\bar{\Lambda}$ mass window &   -   &      0.2    \\
        Wrong combination ratio &        0.8        &   1.4    \\
         MC sample size&   0.4      &   0.5   \\
       Fit range      &        4.5       &   2.6    \\
       Number of type 2 events      &       0.6       &   0.4       \\
       Number of type 3 events      &        0.6      &    0.2      \\
       Signal shape       &     1.7       &   0.6     \\
       Background shape       &      1.4      &   3.5     \\
       Quoted $\phi$ branching fraction&    1.0 &  1.0    \\
       Quoted $\Lambda$ ($\bar{\Lambda})$ branching fraction&    0.8  &  0.8    \\
       Number of $\psi(3686)$ events&     0.5      &    0.5    \\
       \hline
       Individual uncertainty         &   5.2     &   4.7  \\
       Common uncertainty &   2.7  &2.6 \\
       Total &    5.9    &   5.4     \\

\hline
\hline

\end{tabular}
\end{table}

Taking tag A and tag B as two independent measurements, the covariance matrix ($V$) of $\Br_{A}$ and $\Br_{B}$ is

\begin{equation}
\fontsize{8}{10}\selectfont
V=
\begin{bmatrix}     
    ({\sigma}^{A}_{\textrm{sys}}\cdot{\Br_{A}})^{2}+{{{\sigma}^{A}_{\textrm{stat}}}^{2}}  &   ({\sigma}^{A}_{\textrm {sys,\,c}}\cdot{\Br_{A}})({\sigma}^{B}_{\textrm {sys,\,c}}\cdot{\Br_{B}}) \\           ({\sigma}^{A}_{\textrm {sys,\,c}}\cdot{\Br_{A}})({\sigma}^{B}_{\textrm {sys,\,c}}\cdot{\Br_{B}})  &({\sigma}^{B}_{\textrm{sys}}\cdot{\Br_{B}})^{2}+{{{\sigma}^{B}_{\textrm{stat}}}^{2}} \\ 
\end{bmatrix},
\end{equation}
in which $\Br$ denotes the branching fraction of one single tag, ${\sigma}_{\textrm {sys,\,c}}$ denotes the correlated systematic uncertainty, ${\sigma}_{\textrm{sys}}$ denotes the sum of correlated and uncorrelated systematic uncertainties, and ${\sigma}_{\textrm{stat}}$ denotes the statistical uncertainty. The scripts ``$A$" and ``$B$" represent tag A and tag B, respectively. Giving the covariance matrix, the least square method is applied to calculate the weighted average and the uncertainty of the branching fraction, as 

\begin{equation}
\centering
 \fontsize{9}{10}\selectfont
    \begin{aligned}
    \hat{\Br}& ={\left[\sum_{i,\,j=1}^{2} (V^{-1})_{ij}\right]}^{-1} \left[\sum_{i,\,j=1}^{2} (V^{-1})_{ij}\Br_{j} \right] \\
     &=\frac{(V^{-1})_{11}\Br_{A}+(V^{-1})_{12}\Br_{B}+(V^{-1})_{21}\Br_{A}+(V^{-1})_{22}\Br_{B}}{(V^{-1})_{11}+(V^{-1})_{12}+(V^{-1})_{21}+(V^{-1})_{22}}\\
     &\approx 2.64\times 10^{-6}   
    \end{aligned}    
\end{equation}    
and 
\begin{equation}
\centering
\fontsize{9}{10}\selectfont
\begin{aligned}
    \hat{\sigma}_{\Br}^{2}& ={\left[\sum_{i,\,j=1}^{2} (V^{-1})_{ij}\right]}^{-1}  \\
    & = ({(V^{-1})_{11}+(V^{-1})_{12}+(V^{-1})_{21}+(V^{-1})_{22}})^{-1} \\
        \hat{\sigma}_{\Br,\,\textrm{sys}}& =\sqrt{{\hat{\sigma}_{\Br}}^2-{\sigma_{\Br,\,\textrm{stat}}}^2} \\
    & \approx 0.12 \times 10^{-6},
\end{aligned}
\end{equation}    
where $\hat{\Br}$ represents the expected value of the branching fraction, $\hat{\sigma}_{\Br}$ represents the total uncertainty, $\sigma_{\Br,\,\mathrm{stat}}$ represents the statistical uncertainty in Table III and $\hat{\sigma}_{\Br,\,\textrm{sys}}$ represents the overall systematic uncertainty.

	\section{Summary}\label{sec:summary}\vspace{-0.3cm}
 
   Using $(27.12\pm0.14)\times10^{8}$ $\psipp$ events collected with
   the BESIII detector in $2009$, $2012$ and $2021$, the decay
   $\psi(3686)\to\Sigma^{0}\bar{\Sigma}^{0}$$\phi$ is observed for the
   first time with a significance of 7.6$\sigma$. The branching
   fraction of this decay is measured to be $(2.64 \pm 0.32_{\textrm
     {stat}} \pm 0.12_{\textrm {sys}})\times 10^{-6}$. The measured
   branching fraction is consistent with the previous measurement of
   the isospin partner process $\psi(3686) \rightarrow
   \Sigma^{+}\bar{\Sigma}^{-} \phi$~\cite{SSphi}. The principle of
   isospin conservation is upheld within one standard
   deviation. Possible structures in the $\Sigma^{0}\bar{\Sigma}^{0}$
   and $\Sigma^{0}\phi~(\bar{\Sigma}^{0}$$\phi$) invariant mass
   spectra have been investigated. However, with the current
   statistics, no conclusive evidence is observed.

 
\textbf{Acknowledgement}

The BESIII Collaboration thanks the staff of BEPCII and the IHEP computing center for their strong support. This work is supported in part by National Key R\&D Program of China under Contracts Nos. 2020YFA0406300, 2020YFA0406400, 2023YFA1606000; National Natural Science Foundation of China (NSFC) under Contracts Nos. 11635010, 11735014, 11935015, 11935016, 11935018, 12025502, 12035009, 12035013, 12061131003, 12192260, 12192261, 12192262, 12192263, 12192264, 12192265, 12221005, 12225509, 12235017, 12361141819; the Chinese Academy of Sciences (CAS) Large-Scale Scientific Facility Program; the CAS Center for Excellence in Particle Physics (CCEPP); Joint Large-Scale Scientific Facility Funds of the NSFC and CAS under Contract No. U1832207; 100 Talents Program of CAS; The Institute of Nuclear and Particle Physics (INPAC) and Shanghai Key Laboratory for Particle Physics and Cosmology; German Research Foundation DFG under Contracts Nos. 455635585, FOR5327, GRK 2149; Istituto Nazionale di Fisica Nucleare, Italy; Ministry of Development of Turkey under Contract No. DPT2006K-120470; National Research Foundation of Korea under Contract No. NRF-2022R1A2C1092335; National Science and Technology fund of Mongolia; National Science Research and Innovation Fund (NSRF) via the Program Management Unit for Human Resources \& Institutional Development, Research and Innovation of Thailand under Contract No. B16F640076; Polish National Science Centre under Contract No. 2019/35/O/ST2/02907; The Swedish Research Council; U. S. Department of Energy under Contract No. DE-FG02-05ER41374



\end{document}